\documentclass[twoside,leqno,twocolumn]{article}  
\usepackage{ltexpprt} 

\usepackage{amssymb}
\usepackage{amsmath}
\usepackage{algorithm}
\usepackage{algorithmic}
\usepackage{url}
\newcommand{\vect}[1]{\boldsymbol{#1}}

\usepackage{graphicx} 
\usepackage{subfigure}

\newcommand{\rasp}{RaSP}

\newcommand{\trt}{{\em T}}
\newcommand{\bug}{{\em B}}

\begin{document}

\title{\Large Relationship-aware sequential pattern mining}
\author{Nabil Stendardo \thanks{University of Geneva} \\
\and 
Alexandros Kalousis \thanks{University of Geneva}}
\date{}

\maketitle

\begin{abstract} \small\baselineskip=9pt 
Relationship-aware sequential pattern mining is the problem of mining frequent patterns in sequences in which 
the events of a sequence are mutually related by one or more concepts from some respective hierarchical 
taxonomies, based on the type of the events. Additionally events themselves are also described with a certain 
number of taxonomical concepts. We present RaSP an algorithm that is able to mine relationship-aware
patterns over such sequences; RaSP follows a two stage approach. In the first stage it mines for frequent type patterns and 
{\em all} their occurrences within the different sequences. In the second stage it performs hierarchical mining 
where for each frequent type pattern and its occurrences it mines for more specific frequent patterns in the lower 
levels of the taxonomies. We test RaSP on a real world medical application, that provided the inspiration 
for its development, in which we mine for frequent patterns of medical behavior in the antibiotic 
treatment of microbes and show that it has a very good computational performance given the complexity 
of the relationship-aware sequential pattern mining problem.
\end{abstract}


\section{Introduction}
Frequent pattern mining is one of the most popular paradigms in data mining. 
One of the most well studied problem is that of frequent itemset mining in which we
mine for sets of items that appear often together, 
\cite{DBLP:conf/vldb/AgrawalS94,DBLP:conf/sigmod/AgrawalIS93,DBLP:journals/datamine/MannilaT97}.
To address the huge number of patterns that these methods generate, different formulations
of the itemset mining problem have been proposed such as mining for closed or maximal frequent
patterns \cite{DBLP:conf/icdt/PasquierBTL99,DBLP:conf/sigmod/Bayardo98}. Other forms of frequent
pattern mining include mining for frequent sequences in which item order is important, mining 
in the presence of item taxonomies where the items are described in terms of concept hierarchies, 
e.g. GSP \cite{DBLP:conf/vldb/SrikantA95,DBLP:conf/edbt/SrikantA96}. 
In this paper we will focus on sequential pattern mining in the presence of taxonomies where in 
addition the items---events--- within a sequence are related.
In standard frequent sequence mining 
with taxonomies there can be a considerable loss of information in the abstract patterns, namely 
how items relate. For example, a frequent market basket pattern which states that many 
people buy some product A and then buy a second product B one week later, where A and B 
are abstract concepts from the same taxonomy node, is not particularly informative. However
we know much more if we do know that $A==B$, i.e. these persons buy the same product again
regardless of which product is it in the base level. The type of relationships between events that we will 
introduce and mine over do not limit to equality/difference relationships but they can be general discrete 
multi-value relationships, possibly multilevel described also by taxonomies of concepts. We present a novel
algorithm \rasp\ that is able to mine for relationship-aware sequential patterns in the presence of taxonomies
describing the events and their relationships. The algorithm has two stages in the first stage we mine for frequent
patterns of types and {\em all} their occurences within each sequence of some database of sequences. In the second 
stage we refine these type patterns using the taxonomical information of the events and relationships. The rest 
of the paper is structured as follows: in section~\ref{sec:formal.definition} we give a description of the 
relationship-aware sequences over which we will be mining and define the 
problem of relationship-aware sequential pattern mining; in section~\ref{sec:mining} we describe our mining
algorithm, and discuss its computational complexity in section~\ref{sec:complexity}; in section~\ref{sec:experiments}
we explore the comportment of our algorithm on a medical problem in which we look for patterns of medical behavior; 
in section~\ref{sec:related.work} we discuss the related work and we conclude in section~\ref{sec:conclusion}.

\section{Formal Definition}
\label{sec:formal.definition}
A sequence $\mathbf \Sigma$ is an ordered list of $\Sigma_i$ elements which we denote as 
$\mathbf \Sigma=\langle \Sigma_1 \Sigma_2 \dots \Sigma_n \rangle$.
An element, $\Sigma_i$, can be of two kinds either an \textit{event} $e_i$ or a transaction separator denoted by ";"
which denote a passing of time. Thus an example of a $\mathbf \Sigma$ sequence would be:
\begin{eqnarray}
\label{eq:seq}
{\mathbf \Sigma} = \langle e_{1}e_{2}\dots e_{k};e_{k+2}\dots;\dots e_{n} \rangle 
\end{eqnarray}
By $\mathbf \Sigma^{(e)}$ we denote the sequence that consists only of the events of $\mathbf \Sigma$, i.e. no transaction
separators. The operator $|\mathbf \Sigma|$ denotes the length of the sequence, i.e. the number of events in $\mathbf \Sigma$ plus the
number of transaction operators; the $i$th symbol of the sequence is given by $\Sigma_i$. Similarly the  operator $|\mathbf \Sigma^{(e)}|$ 
gives the number of events of the $\mathbf \Sigma$ sequence and we can retrieve its $i$th event by $\Sigma_i^{(e)}$.
Additionally we define the function $E(\mathbf \Sigma,i)$ which returns the index of the $i$th event in the $\mathbf \Sigma$ sequence, 
i.e. $\mathbf \Sigma_{E(\mathbf \Sigma,i)} = \Sigma_i^{(e)}$, note that $i$ and $E(\mathbf \Sigma,i)$ in general are not
the same due to the presence of transaction separators in the sequence.

An \textit{event} $e$ has some \textit{event type} $t$, $t=T(e)$.
Each event type $t$ has 
an associated array of taxonomies $\mathbf X_t = [X_{t_1},\dots,X_{t_k}]$ that we call the 
{\em event type schema}.
A taxonomy $X$ is a directed tree representing {\em is-a} relationships defined over the set of \textit{concepts} 
$\mathcal C(X)$ that are the nodes of the tree. A directed edge from a concept $c_i$ to a concept $c_j$ denotes that
$c_i$ is a generalization of $c_j$. We say that concept $c_i$ {\em subsumes} $c_j$, denoted by $c_i \models c_j$, 
if there is a directed path in the tree starting from $c_i$ and arriving at $c_j$.
Additionally for every pair of event types, $t_i,t_j$, we have a symmetric relationship  which has a {\em relationship type}
$\rho_{t_it_j}$. Similarly to the event types any relationship type is associated with a array of taxonomies  
$\mathbf X_{\rho_{t_it_j}}=[X_{\rho_{{t_it_j}_1}},\dots,X_{\rho_{{t_it_j}_d}}]$ that we call the {\em relationship type 
schema}.  Note that that event type schemas and relationship type schemas can be empty.
If the length of the relational type schema is zero for all relationship types we are in the framework of 
frequent sequential pattern mining in the presence of taxonomies, e.g. GSP, if in addition the event types
have a zero length event type schema then we are in the standard frequent sequential pattern mining framework.
It is possible to include also n-ary relations however for simplicity we omit that. 

An event $e$ of type $t$ has an associated {\em event-concepts array} $\mathbf c(e)=[c_1, c_2, \dots c_k]$,
of the same length as the taxonomies array of its type, where $c_i \in \mathcal C(X_{t_i})$; note here that
events of the same type can be associated to concepts that are found on different abstraction levels of the
respective ontologies. 
Similarly for every pair of events, $e_k, e_l$, in some sequence, with corresponding event types $t_i$ and $t_j$ which define
the respective relationship type $\rho_{t_it_j}$, we have an associated {\em relationship-concepts array} 
$\mathbf r(e_k,e_l)=[r_1, r_2, \dots r_d]$, of the same length as the relationship type schema of $\rho_{t_it_j}$, 
where $r_i \in \mathcal C(X_{\rho_{{t_it_j}_i}})$.
The subsumption operator is also defined for arrays of concepts provided that these have the same 
length and their $k$th elements belong to the same ontology $X_k$. For any two concept arrays $\mathbf c_i, \mathbf c_j$,
we will say that $\mathbf c_i \models \mathbf c_j$ if $\forall k, c_{i_k} \models c_{j_k}$. The sequences 
that we will be considering in this paper will be {\em relationship-aware sequences} that are fully described by the events 
they contain and the relationships between these events as discussed above.



A {\em transaction} is the sequence of events betweem two consecutive transaction separators of a $\mathbf \Sigma$
sequence, or the complete sequence from the beggining of $\mathbf \Sigma$ to the first transaction separator.
Events in a transaction are assumed to be different.  They are sorted first by the lexicographical 
order of their respective types, and then, in case of type equality, by their concept arrays according to the first 
different concept. The concept order is given by a pre- or post-order depth first traversal of the respective
taxonomy. This ensures that two semantically identical transactions have the same transcription. We assume that no
empty transactions exist.

We will define a number of sequence representations that will be used by our algorithm.
We will construct the {\em type-and-concept-aware} representation, $\mathbf \Sigma_{a}$, of a sequence $\mathbf \Sigma$ in a manner 
that will contain for each of its events, $e_k \in \mathbf \Sigma$, their type, $T(e_k)$, and event concept arrays, $\mathbf c(e_k)$,
and for each pair of its events, $e_k,e_l$, their respective relationship concept arrays, $\mathbf r(e_k,e_l)$. Since relationships are symmetric 
we only need to include one of the two $\mathbf r(e_k,e_l)$, $\mathbf r(e_k,e_l)$, relationship concept arrays in the concept
aware representation. We will include these $\mathbf r(e_k,e_l)$ for which $l < k$, i.e. we will describe the relationships of
every event only to its preceding events, and for each $k$ we will order the $\mathbf r(e_k,e_l)$ in ascending order from $l=1$ to 
$l=k-1$. Thus the representation of each event $e_k$ in a sequence will now be a complex structure of the form
$$\bar{e_k} = [T(e_k), \mathbf c(e_k), \mathbf r(e_k,e_1), \mathbf r(e_k,e_2), \dots, \mathbf r(e_k,e_{k-1})]$$
in which the first element gives the type of the event, the second is the concept array of the event, and the remaining ones
are the different relationship concept array of this event with all its previous ones in the sequence. 
The type-and-concept-aware representation of the example sequence given in eq~\ref{eq:seq} will be:
$\mathbf \Sigma_{a} = \langle \bar{e_{1}}\bar{e_{2}}\dots \bar{e_{k}};\bar{e_{k+1}}\dots;\dots \bar{e_{n}} \rangle$. In addition 
to the type-and-concept-aware representation of a sequence we will also define its {\em type-aware} and {\em concept-aware} 
representations. 
The former will be the sequence that only contains the type information of the events and the latter the 
sequence constructed from the concatenation of the elements of the events' event-concepts and relationship-concepts arrays. 
For the example sequence given in eq~\ref{eq:seq} the type-aware representation will be:
$$\mathbf \Sigma_{ta}= \langle T(e_{1})T(e_{2})\dots T(e_{k});T(e_{k+1})\dots;\dots T(e_{n}) \rangle $$ 
and the concept-aware: 
\begin{eqnarray*}
\mathbf \Sigma_{ca} & = & \langle \mathbf {\hat c}(e_1) \mathbf {\hat c}(e_2) 
                          \mathbf {\hat r}(e_2,e_1) \dots \mathbf {\hat c}(e_k) \mathbf {\hat r}(e_k,e_1) \dots \mathbf{\hat r}(e_k,e_{k-1}) \\
                    &   & \mathbf {\hat c}(e_{k+1})  \mathbf {\hat r}(e_k,e_1) \dots \mathbf{\hat r}(e_k,e_{k})  \\
                    &   & \dots \mathbf {\hat c}(e_{n})  \mathbf {\hat r}(e_k,e_1) \dots \mathbf {\hat r}(e_n,e_{n-1})\rangle 
\end{eqnarray*}
where for an array $\mathbf v=[v_1,v_2,\dots,v_k]$, $\mathbf{\hat v}$ returns the elements of $\mathbf v$ in the order in which they appear in $\mathbf v$ 
i.e. $\mathbf {\hat v} = v_1 v_2\dots v_k$.

The projection of a sequence $\mathbf \Sigma$ to an array of indeces, $\mathbf v$, is that subsequence whose 
elements are the elements of the original sequence whose indeces are given by $\mathbf v$. We will denote the projection of
$\mathbf \Sigma$ onto $\mathbf v$ by $\mathbf \Sigma \perp \mathbf v$. Obviously since a projection of a sequence is also a
sequence we can have its different representations introduced previously.

A {\em relationship-aware sequential pattern}, $\mathbf \Pi$, is defined in the same manner 
as a relational sequence with the difference that, unlike sequences, it can have identical 
elements within transactions. 
Similarly to the sequences notation $|\mathbf \Pi|$ denotes the length of the pattern, i.e. number of events plus the
number of transaction operators, $\Pi_i$ is the $i$-th element of the pattern, $\mathbf \Pi^{(e)}$ is the pattern that consists
of only the events of $\mathbf \Pi$, i.e. no transaction separators, $\mathbf \Pi^{(e)}_i$ is its $i$th element, and $|\mathbf \Pi^{(e)}|$ 
is the number of events of the $\mathbf \Pi$ pattern. Moreover patterns also have the different types of representations that we
presented for sequences, i.e. type-and-concept-aware, type-aware and concept-aware.
Some additional comments on notation are also in order. In general we will denote 
arrays by bold variables, e.g. $\mathbf v$,
and the $i$th element of an array with a normal typeset, e.g. $v_k$; the notation $\mathbf v_k$ will be used to index arrays and not
their elements. For brievity reasons we will refer to relationship-aware patterns as patterns.

A pattern $\mathbf \Pi$ {\em matches} to a sequence $\mathbf \Sigma$ if and only if there exists an {\em events index vector}
$\vect \lambda=(\lambda_1, \lambda_2, \dots, \lambda_{|\mathbf \Pi^e|})$ in $\mathbf \Sigma^{(e)}$, 
whose length is equal to the number of events in  $\mathbf \Pi$, and $1 \leq \lambda_i \leq |\mathbf \Sigma^{(e)}|$,
such that: 
\begin{itemize}
\item $\forall \Pi_{i}^{(e)} \Rightarrow T(\Pi_i^{(e)}) == T(\Sigma^{(e)}_{\lambda_{i}})\ \wedge\ 
                                                           \mathbf c(\Pi^{(e)}_i)\models \mathbf c(\Sigma^{(e)}_{\lambda_{i}})$ 
\item $\forall (\Pi_{i}^{(e)},\Pi_{j}^{(e)})\ \wedge i\neq j\Rightarrow \mathbf r(\Pi_{i}^{(e)},\Pi_{j}^{(e)})\models 
                \mathbf r(\Sigma^{(e)}_{\lambda_{i}},\Sigma^{(e)}_{\lambda_{j}})$ 
\item Any pair of events in $\mathbf \Sigma^{(e)} \perp {\vect \lambda}$ which are separated by at least one transaction 
separator in the $\mathbf \Sigma$ sequence must also be so in $\mathbf \Pi$, and vice-versa, i.e. $a;b$ does not match 
to $ab$, nor the opposite. 
\end{itemize}
Note that to get the events index vector with respect to the full $\mathbf \Sigma$ sequence we just need to apply 
element-wise the function $E(\mathbf \Sigma, )$ on the $\vect \lambda$ vector; we will denote this element-wise application
and the resulting index vector by $E(\mathbf \Sigma,\vect \lambda)$. 


Additionally we define a number of constraints on the form of the
mined patterns namely a {\em max-gap} constraint, $mg$, and a {\em maximum-projected-length} constraint, $mpl$, by defining
constraints on the respective indeces vectors $\vect \lambda$. The max-gap constraint is defined as 
$\lambda_i - \lambda_{i+i} \leq mg$ and the maximum-projected-length constraint as $\lambda_{|\vect \lambda|} - \lambda_1 \leq mpl$.

Given a database of sequences the {\em support} of a pattern is the number of sequences it matches. 
We will call an {\em occurence} of a pattern $\mathbf \Pi_l$ the couple $O_{l_k}=(\mathbf \Sigma_{l_k}, \vect \lambda_{l_k})$
where $\mathbf \Sigma_{l_k}$ is some sequence and $\vect \lambda_{l_k}$ is the event index vector of that pattern
in $\mathbf \Sigma^{(e)}_{l_k}$. Each pattern is associated with a set of occurences $\mathbf O_l=\{O_{l_k} | k = 1 \dots M \}$.
The cardinality, $|\mathbf O_l|$, of the set of occurences is the {\em number of occurences} of the $\mathbf \Pi_l$ 
pattern. Clearly a given sequence can include multiple repetitions of the same pattern, thus it can be that
for different occurences $O_{l_i}$ and $O_{l_j}$ of the $\mathbf \Pi_l$ pattern that 
$\mathbf \Sigma_{l_i}==\mathbf \Sigma_{l_j}$ in which case nevertheless the respective vector of 
indeces $\vect \lambda_{l_i}$ and $\vect \lambda_{l_j}$ will be different. The number of occurences 
of a pattern is trivially larger or equal to its support.

Given a set of relational sequences $\mathbf S=\{\mathbf \Sigma_1, \mathbf \Sigma_2, \dots, \Sigma_{|\mathbf S|}\}$ 
we will define as {\em Relationship-aware Sequential Pattern Mining} the discovery of all relationship-aware patterns, 
or a well-defined subset of them, with support in $\mathbf S$ larger than some threshold $\theta$. In the next section 
we will describe an algorithm, RaSP, for mining such patterns.

\section{Mining for relationship-aware sequential patterns} 
\label{sec:mining}
We address the problem of mining for relationship-aware sequential patterns using a two stage algorithm. 
In the first stage we will mine for frequent patterns of types, i.e. we will be mining over the 
type-aware, $\mathbf \Sigma_{ta}$, representations of the sequences.  We will call such a resulting pattern a 
{\em type-pattern} and denote it by $\mathbf \Pi_{{ta}}$; we use the subscript $\ _{ta}$ to emphasize that these patterns are 
in the type-aware representation, like the sequences from which they were produced, since they are type-patterns.
For each one of these patterns we will compute its associated set of occurences. Mining over the type-aware 
representation of the sequences gives us patterns that are defined only over the roots of the event-type and 
relationship-type taxonomies. To get the final refined\footnote{The refined patterns will be defined over 
the different levels of the taxonomies and not just their roots.} frequent patterns of each frequent type-pattern $\mathbf \Pi_{{ta}}$
in the second stage of the algorithm we retrieve for each $\mathbf \Pi_{{ta}}$ and its set of occurences all 
the associated sequences. Based on the concept-aware representation 
of these sequences we get the refined patterns by solving one frequent pattern mining problem {\em for each} frequent 
type-pattern. 
We should note here that given a specific type-pattern the associated concept-aware sequences are all of the same 
length. 
A basic difference of our approach from standard frequent pattern mining algorithms is that 
we need to compute all the occurences of a given frequent type-pattern, standard pattern mining algorithms only look at
the presence (typically the first) or absence of a pattern in a sequence.
We need the occurences of each frequent type-pattern because it is over them that we can compute the 
different frequent refinements of a given type-pattern. If we were limited only to the first presence 
of a type-pattern in a sequence we would be computing a distorted and incomplete picture of its refined 
patterns.

\subsection{Mining for type-patterns}
In this stage of the algorithm we will be mining for frequent type-patterns on the sequence 
set $\mathbf S$  over the type-aware representation of the sequences. We cannot use a standard off-the-self frequent pattern mining algorithm since 
as just mentioned these do not discover all occurences within a sequence of a given pattern. 
In order to discover all of them we need to modify 
some frequent pattern mining algorithm. 
We have opted for a modification of the well known GSP algorithm, {\cite{DBLP:conf/edbt/SrikantA96}, because
it was easier to adapt it to the requirements of our problem 
compared to more efficient alternatives such as PrefixSpan, \cite{DBLP:conf/icde/PeiHPCDH01}.

The GSP algorithm is based on the Apriori property which states that if a pattern $\mathbf \Pi$ is frequent then 
all its subpatterns are also frequent. GSP works in an iterative manner it first finds all frequent patterns of 
length $k$ and then given these patterns it determines all possible candidate patterns of length $k+1$ according 
to the Apriori property, {\em candidate generation} phase, and subsequently checks these patterns against the 
database to determine their frequency, {\em candidate checking} phase. To adapt GSP so that it also computes the
set of occurences of each pattern we only need to significantly modify the candidate checking phase, the 
candidate generation hardly changes. 

In Algorithm~\ref{alg:Simple-Candidate-Checking} we give a basic algorithm that finds {\em all} occurrences of 
a $\mathbf \Pi$ pattern in a $\mathbf \Sigma$ sequence with no transaction separators. The main functionality
of the algorithm is delivered by the {\bf MGSP\_SCC} function.
The basic idea is that given a element match, $ \Pi_{i}==\Sigma_{j}$, 
we continue the search with $i\leftarrow i+1$ and $j\leftarrow j+1$, i.e. we account for the specific element match, but also with only $j=j+1$ while 
keeping $i$ unchanged, i.e. we ignore this match. If there is no match, we just increment $j$ and continue. In Algorithm~\ref{alg:General-Candidate-Checking}
we give the more evolved function, {\bf MGSP\_GCC}, for candidate checking and all occurences finding for patterns and sequences that contain transaction 
separators which is follows the same lines as its simpler version {\bf MGSP\_SCC}. The utility function next($\mathbf Z$,i) indicates the next element
in the sequence or pattern $\mathbf Z$ which is not a transaction separator. 

We implemented certain optimizations to speed up the algorithm. For
each type-pattern to check, as well as for each sequence in our database,
we create a multiset of the elements in the form of a vector. For
instance, given types a,b,c and d, the vector $m_{\mathbf X}$(where $\mathbf X$ is
a sequence or a pattern) is equal to {[}\#a $\in \mathbf X$, \#b $\in \mathbf X$, \#c $\in
\mathbf X$,\#d $\in \mathbf X${]}, where {}``\#$x$ $\in \mathbf X$'' is the number of times 
type $x$ appears in the sequence or pattern $\mathbf X$. When checking type-pattern
$\vect \Pi$ against sequence $\vect \Sigma_{ta}$, we first compute the vector $m_{d}=m_{\vect \Sigma}-m_{\vect \Pi}$,
and if any element is negative, we return immediately (no match). We
also pass as parameter to the algorithm a local $m_{L}$ vector which
is initialized at the first call to $m_{d}$, and which is incremented
by 1 in position $i$ any time one advances over $t_{i}$ in $\vect \Pi$,
and decremented at position $i$ any time one advances over $t_{i}$
in $\vect \Sigma+{ta}$. If while advancing over a sequence element, that element
becomes negative, we have an impossibility to find a match in
the future, and we return immediately. We also store, for each candidate 
of length $k$, the sequence identifiers in which all its parent frequent patterns 
of length $k-1$ have matched, and test only those sequences.

The max-gap constraint is trivial to implement, it is a simple check 
and return, and so is the maximum-projected-length constraint.
However one needs to be careful in defining constraints so that the Apriori
property is respected. Concretely if the occurrence of a pattern of 
length $k$ passes the test, all possible $k$ combinations
of length $k-1$ must also pass the test. An example of invalid constraint
is: {}``We want at least 2 different event types in a pattern''
because it cannot be satisfied at level one, but can be satisfied at
higher levels.

\begin{algorithm}
\caption{\label{alg:Simple-Candidate-Checking}Modified GSP Simple Candidate Checking}
\begin{algorithmic}
\small
\STATE \COMMENT{$\mathbf \Pi$ A pattern}
\STATE \COMMENT{$\mathbf \Sigma$ The sequence within which we look for the occurence of $\mathbf \Pi$}
\STATE $\mathbf O=\emptyset$ \COMMENT{The set of occurences of pattern $\mathbf \Pi$ in sequence $\mathbf \Sigma$}
\STATE call MGSP\_SCC($\mathbf \Sigma$,$\mathbf \Pi$)
\STATE
\STATE {\bf function}~MGSP\_SCC($\mathbf \Sigma$,$\mathbf \Pi$,i=1,j=1,$\vect \lambda=[]$) \\
\COMMENT {$i$ and $j$ are indeces in the $\mathbf \Pi$ pattern and $\mathbf \Sigma$ sequence respectively; 
$\vect \lambda$ is the index vector of the $\mathbf \Pi$ pattern in the $\mathbf \Sigma$ sequence.}
\IF{i $>$ $|\mathbf \Pi|$}
 \STATE $\mathbf O = \mathbf O \cup \mathbf O \{ (\mathbf \Sigma, \vect \lambda) \}$
 \STATE return
\ENDIF
\IF{j $>$ $|\mathbf \Sigma|$}
 \STATE return \COMMENT{No~match}
\ENDIF
\IF{$\Pi_i$==$\Sigma_{j}$}
 \STATE MSGP\_SCC($\mathbf \Sigma$,$\mathbf \Pi$,i+1,j+1,[$\vect \lambda$,j])
\ENDIF
\STATE MSGP\_SCC($\mathbf \Sigma$,$\mathbf \Pi$,i,j+1,$\vect \lambda$)
\end{algorithmic}
\end{algorithm}

\subsection{From frequent type-patterns to frequent relational patterns}
At the end of the first stage of the algorithm we have a set $\mathbf P = \{\mathbf \Pi_{l_{ta}} | l=1\dots L\}$ of type-patterns, where each 
$\mathbf \Pi_{l_{ta}}$ type-pattern is associated with a set of occurences $\mathbf O_l=\{O_{l_k} = (\mathbf \Sigma_{l_k}, \vect \lambda_{l_k})| 
k = 1\dots {M}_l\}$. 
In the second stage of the algorithm we will define a frequent itemset mining problem for each $\mathbf \Pi_{l_{ta}}$ 
type-pattern on the basis of its associated occurence set, $\mathbf O_l$, thus we will be solving $|\mathbf P|$ different 
frequent itemset mining problems. The patterns that will be computed for each of these itemset mining problems will give 
rise to the final refined frequent patterns of the original type-pattern.

Given an occurence $O_{l_k}$ we define $\mathbf u_{l_k}$ to be the projection of the $\mathbf \Sigma^{(e)}_{l_k}$ event sequence on the
events index vector $\vect \lambda_{l_k}$, i.e. $\mathbf u_{l_k} = \mathbf \Sigma^{(e)}_{l_k} \perp \vect \lambda_{l_k}$, 
which is also equivalent to the $\mathbf \Sigma_{l_k} \perp E(\mathbf \Sigma_{l_k} ,\vect \lambda_{l_k})$ projection in 
terms of the full sequence. Since all $\mathbf u_{l_k}$ have been produced from the same type-pattern $\mathbf \Pi_{l_{ta}}$ they 
all have the same number of elements which is equal to the number of events in  $\mathbf \Pi_{l_{ta}}$, i.e.  $|\mathbf \Pi^{(e)}_l|$.
We will denote with $\mathbf u_{l_{k_{ca}}}$ the concept-aware representation of $\mathbf u_{l_k}$. 
Remember that the concept-aware representation
contains both the event and relationship concepts. Let $c_j$ be the $j$th element (concept) of $\mathbf u_{l_{k_{ca}}}$. Since $c_j$ 
is a concept it is associated with some ontology $X$ and we denote by $A(c_j,X)$  the set that contains all its ancestors in its taxonomy, 
including itself if not the root but excluding the root. We denote by $\mathbf \Gamma_{l_{k_{ca}j}} = \{ (a, j) | a \in A(c_j,X)\}$ 
the set that describes for the projection sequence $\mathbf u_{l_{k_{ca}}}$ which are the ancestor concepts of the concept $c_j$ that 
appears in its $j$th position coupled with their position information $j$.  We define the {\em concept-position set}, $\mathbf \Gamma_{l_{k_{ca}}}$,
of the projected sequence $\mathbf u_{l_{k_{ca}}}$ as $\mathbf \Gamma_{l_{k_{ca}}}=\cup_j \mathbf \Gamma_{l_{k_{ca}j}}$
that gives which taxonomies concepts appear in which positions of $\mathbf u_{l_{k_{ca}}}$. Subsequently we construct the $\vect \Omega_l$ 
set which contains the union of the concept-position sets over all the occurences of the $\mathbf O_l$ set,  
$\vect \Omega_l=\cup_k \mathbf \Gamma_{l_{k_{ca}}}$, which we index based on some lexicographical order of its items. We can think of
the $\vect \Omega_l$ set as a vocabulary that is created of all couples of the form $(some\ ancestor\ concept\ 
of\ c_j,j | c_j \in \mathbf u_{l_{k_{ca}}})$ 
gathered from the projections $\mathbf u_{l_{k_{ca}}}$ that are produced from all occurences of the $\mathbf \Pi_{l_{ta}}$  pattern.
We will use this vocabulary to describe the different $O_{l_k}$ occurences of the set of occurences $\mathbf O_l$ of $\mathbf \Pi_{l_{ta}}$ 
by defining the matrix $\mathbf O'_l: M_l \times |\vect \Omega_l|$ whose $k$th row corresponds to the $O_{l_k}$ occurence of $\mathbf O_l$
and the columns to the different elements of the $\vect \Omega_l$ vocabulary. The $(k,h)$ element of the $\mathbf O'_l$ matrix will be 
one if the $O_{l_k}$ occurence contains in its concept-position set $\mathbf \Gamma_{l_{k_{ca}}}$ the $(some\ ancestor\ concept\ of\ c_j,j)$ 
pair that corresponds to the $h$th element of  $\vect \Omega_l$. 

It is on the  $\mathbf O'_l$ matrix representation of the set of occurences $\mathbf O_l$ that the frequent itemset mining will take 
place. We define an itemset $\mathbf i$ on  $\mathbf O'_l$ as a vector of column indeces, i.e. $\mathbf i=(i_1, i_2,\dots, i_{\alpha}), 
\alpha \leq |\vect \Omega_l|$. 
It is important to note that an itemset $\mathbf i$ actually corresponds to a highly redundant refinement of the $\mathbf \Pi_{l_{ta}}$
type-pattern that contains events and relationships concepts from different levels of the associated taxonomies ($\mathbf \Pi_{l_{ta}}$
contains only event types and implicitly relationship types; highly redundant because it includes all anscestor concepts
of a concept that appears in some position j). In fact from these itemsets we will construct the final relationship-aware 
patterns we are mining for. 
We can easily get the subset of occurences, $\mathbf O_{l_{\mathbf i}} 
\subseteq \mathbf O_l$, that contain a given itemset $\mathbf i$ through a column vector, $\vect \mu_{\mathbf i}$, that is created by the pairwise 
multiplication of the columns of $\mathbf O_l$ that are 
indexed by $\mathbf i$. If $\mu_{\mathbf i_k} ==1$ then the $\mathbf O_{l_k}$ occurence contains the $\mathbf i$ itemset. However unlike standard frequent
itemset mining we do not measure the support of the $\mathbf i$ itemset on the occurence set $\vect \Omega_l$ but on full  set of sequences 
$\mathbf S$ since it is the support in this set that we are interested in. To do so we need one additional matrix $\vect \Phi: M_l \times 
|\mathbf S|$ whose rows correspond again to the occurences in $\vect \Omega_l$ and the columns to the different sequences of $\mathbf S$, 
this matrix simply indicates to which sequence a given occurence belongs (all elements of a given row are zero except that which
corresponds to the sequence to which the occurence belongs which is set to one, since each occurence appears in exactly one sequence). 
So the support of the $\mathbf i$ itemset will be given by the number of columns (i.e. number of sequences) 
of the $\vect \Phi$ matrix that have at least one common non-zero entry with the $\vect \mu_{\mathbf i}$ column vector.
We will call the problem of finding all $\mathbf i$ itemsets, or a well defined subset of them such as closed or maximal 
itemsets\footnote{A maximal itemset is a frequent itemset that has no super set that is frequent, i.e. all its super sets 
have support lower then $\theta$; a frequent itemset is closed if there is no super set of it with the same support.
}, 
whose support in $\mathbf S$ is larger than some $\theta$ threshold an occurence/itemset mining problem. It can be solved by simple adaptations 
of standard frequent itemset mining problems, e.g. Apriori. However frequent itemset mining algorithms that are based on properties which are 
not valid in our case, such as $sup(S) = sup(S \cup \{a\}) = sup(S \cup \{a\}) \Rightarrow sup(S \cup \{a, b\}) $, require more care in their adaptation.
We used a simple adaptation of GenMax~\cite{DBLP:journals/datamine/GoudaZ05} (we did not use diffsets), which finds maximal itemsets.

Finally from the $\mathbf i$ itemset we will now construct the corresponding concept-aware representation, $\mathbf \Pi^{(\mathbf i)}_{l_{ca}}$,  
of the $\mathbf \Pi_{l_{ta}}$ type-pattern. 
Remember that the itemset is a set of the form $\mathbf i= \{  (some\ ancestor\ $ $concept\ $ $of\ c_j,j)$ $|  $ $c_j \in \mathbf u_{l_{k_{ca}}}, $ $
\mathbf u_{l_{k_{ca}}}\ derived\ from\ some\ O_{l_k}\}$. The concept in the $j$ position of the $\mathbf \Pi^{(\mathbf i)}_{l_{ca}}$ representation 
will be, either the taxonomy root  if there exists no element in $\mathbf i$ which contains the $j$ index,
or it will be $c$ if $(c,j) \in \mathbf i$ and there is no descendant $d$ of $c$ such that $(d,j) \in \mathbf 
i$, i.e. we include the most specific concept its time. Since 
we cannot have in some given position $j$ two concepts which are unrelated to each other by a descendance relationship,
because of the tree structure of the taxonomies and our definition of the occurence/itemset mining problem, the element 
in the position $j$ will be unique. We can extend our approach to include generalization relationships that are defined
over Directed Acyclic Graphs instead of tree-like taxonomies in which case we would have multiple elements in the same 
position. One way to address this is by looking for a common ancestor of these concepts which assumes that we have a 
rooted DAG, yet another one would be to extend our pattern representation to allow conjunction of concepts. 

From the concept-aware represenation $\mathbf \Pi^{(\mathbf i)}_{l_{ca}}$ and its $\mathbf \Pi_{l_{ta}}$ 
type-pattern we can have the final pattern representation $\mathbf \Pi^{(\mathbf i)}_{l}$.
So finally for each type-pattern $\mathbf \Pi_{l_{ta}}$ we will have a set, $\mathbf R_l$, of frequent relationship-aware 
patterns $\mathbf R_l=\{ \mathbf \Pi^{(\mathbf i)}_{l_{ca}} | sup(\mathbf i) \geq \theta \}$, and the complete set, $\mathbf R$, of frequent
relationship-aware patterns will be given by the union of these sets, $\mathbf R = \cup_l \mathbf R_l$.

\begin{algorithm}
\caption{\label{alg:General-Candidate-Checking}Modified GSP General Candidate Checking}
\begin{algorithmic}
\small
\STATE {\bf function}~MGSP\_GCC($\mathbf \Sigma$,$\mathbf \Pi$,i=1,j=1,$\vect \lambda$=[],\\~~~~~~~~~~~~~~~~~~~~~~~~~~~~~ lastMatchWasTS=True,TS=";")
\IF{i $>$ $|\mathbf \Pi|$}
 \STATE $\mathbf O = \mathbf O \cup \mathbf O \{ (\mathbf \Sigma, \vect \lambda) \}$
 \STATE return
\ENDIF
\IF{j~>~$|\mathbf \Sigma|$}
 \STATE return \COMMENT{No~match}
\ENDIF
\STATE \COMMENT{Optional: Do check for multiset inclusion}
\STATE \COMMENT{Optional: Do check for max-gap violation}
\STATE \COMMENT{Optional: Do check for max-projected-length violation}
\STATE def~previousInSequenceIsTS=($\Sigma_{j-1}==TS$)
\STATE def~nextInPatternIsTS=($\Pi_{i+1}==TS$)
\IF{previousInSequenceIsTS \&\& !lastMatchWasTS}
 \STATE return \COMMENT{Found a TS in sequence which was not in pattern.}
\ENDIF
\IF{$\Pi_{i}$~==~$\Sigma_{j}$}
 \IF{nextInPatternIsTS}
   \STATE   nextPos~=~position~of~the~item~just~after~the~next~TS~in~$\mathbf \Sigma$
   \IF{nextPos==NULL}
     \STATE return \COMMENT{No~match}
   \ENDIF
     \STATE MSGP\_GCC($\mathbf \Sigma$,$\mathbf \Pi$,next($\mathbf \Pi$,i), nextPos,$[\vect \lambda,j]$,True)
   \ELSE
     \STATE MSGP\_GCC($\mathbf \Sigma$,$\mathbf \Pi$,next($\mathbf \Pi$,i), next($\mathbf \Sigma$,j),$[\vect \lambda,j]$,False)
   \ENDIF
 \ENDIF
\STATE MSGP\_GCC($\mathbf \Sigma$,$\mathbf \Pi$,i,j+1, $\vect \lambda$,lastMatchWasTS)
\end{algorithmic}
\end{algorithm}

\section{Computational Complexity}
\label{sec:complexity}
The computational complexity of the second stage of the algorithm is 
determined by the GenMax algorithm that we use there and is a multiple
of the number of the type-patterns that were discovered in the 
type-pattern mining stage. The type-pattern mining complexity, assuming we are using the modified
GSP described in this paper, however, has an easily-determinable upper
bound. In the case of matching a pattern of size $k$ to a sequence
of size $n>k$, there are at most $\left(\begin{array}{c}
n\\
k\end{array}\right)$possible occurrences to find, and so the theoretical worst-case complexity
is $O\left(\left(\begin{array}{c}
n\\
k\end{array}\right)k\right)$, which is much larger than the $O(n)$ needed to determine the
presence or the absence of a pattern in a sequence. However, with
a max-gap constraint of $g$ events, an upper bound on the number
of occurrences becomes $(g-1)^{k-1}(n-k)$ so an upper bound in the
complexity is $O\left((g-1)^{k-1}(n-k)k\right)$ instead of the presence-only
$O(n^{2})$ forward-backward described in \cite{DBLP:conf/edbt/SrikantA96}.
If we include a max projected length (in terms of elements) $w$ $(n>w>k)$,
an upper bound on the number of occurrences becomes $\left(\begin{array}{c}
w\\
k-1\end{array}\right)(n-k)$ and so an upper bound in complexity is $O\left(\left(\begin{array}{c}
w\\
k-1\end{array}\right)(n-k)k\right)$.
Even if this worst-case computational complexity may seem overwhelming,
it only actually happens when the type-pattern $\vect \Pi$ and the type-aware sequence representation
$\vect \Sigma_{ta}$ are equal to $T^{k}$ and $T^{n}$ (where $T$ is a single
event type) respectively. 
However, computational complexity and combinatorial explosion of occurrences can 
be a problem in some cases. Possible remedies are:
adding a max-gap and/or max projected length constraint, either in
terms of event count or in terms of time; artificially disregarding 
type-patterns consisting of $k$ times the same event type in a row (and all of its children);
removing outliers in the dataset, such as extraordinarily long sequences, or sequences with 
more than $q$ events of the same type.

Another problem of our approach can be memory usage. It is true that
storing all occurrences in main memory can become prohibitive. A solution
to that problem can be to stream the occurrence data of a type-pattern
to a file (or similar linear storage), and then release the main
memory associated with it. This should be done once all occurrences
of a type-pattern are found and has been determined as frequent.
In addition to avoid memory saturation while solving an occurrence/itemset
problem, a few solutions are available. We can use a sparse representation of occurrence vectors (our $\mu$-values).
Alternatively we can keeping only $k$ occurrences per sequence when the total number of occurrences
is higher than a value $n$. This makes the algorithm incomplete, but in the case of a small number 
of outlying sequences, it does not significantly impact the usefulness of our method, while it significantly improves performance.

\section{Experiments}
\label{sec:experiments}
We will test various aspects of the performance of our method on a dataset that is 
associated to a medical problem. In fact this medical problem was the trigger that 
led to the development of \rasp. The general problem that we want to address is the 
extraction of patterns of medical practice in what has to do with the antibiotic 
treatment of microbes. We have a dataset that contains 6659 episodes of care which 
have at least one antibiotic treatment which span nine years from 2002 to 2010. 
The data have been collected from the clinical sites that participate in European project DebugIT~\cite{debugIT}.
An episode of care is the sequence of events that take place during the stay of 
some patient in the hospital. Since we are interested 
in the extraction of antibiotic treatment patterns the types of events that we are interested 
in are two. The first event type is \bug\ which corresponds to the detection---presence--- of a microbe at some moment 
in time during the episode of care, the detection is done through a laboratory test; the same
microbe can be detected more than once within the course of a given episode of care and each detection
is considered as a different event of the \bug\ type. The second event type is \trt\, 
which corresponds to the prescription of a treatment with a given drug at some moment during the 
episode of care. In addition we define the following relationship types \bug\ $\times$ \trt, \bug\ $\times$ \bug, and \trt\ $\times$ \trt. 
The first relation corresponds to the notion of an antibiogram. An antibiogram is a laboratory
test in which we test the sensitivity of a detected bug on some drug; if it takes place it does 
so right away after the detection of the bug. The two latter relations
are simply identity relations which indicate whether their respective arguments, i.e. events,
are the same or different.

With each event type, \bug\ and \trt, there is one taxonomy associated, i.e. the respective event-type schemata
have a length of one. With the events of type microbe the associated taxonomy is NewT from Uniprot, and more precisely 
that part of it which has to do with the bacteria-microbes, \cite{NewT}. With the events of type \trt\ the associated taxonomy is the
Anatomical Therapeutic Chemical Classification System, ATC, which is used for the classification of drugs, and more precisely 
that part of the taxonomy which is related to Antibiotics (antibacterials for systemic use, with associated ATC code starting 
with J01), \cite{ATC}. In figure~\ref{fig:taxo} we give a snapshot of the two taxonomies. Some additional statistics on these
two taxonomies are given in table~\ref{table:taxo}.
The three relationships are also associated with one taxonomy each; here the associated taxonomies 
are simpler. The taxonomy associated with \bug\ $\times$ \trt\ is given by the following prefix tree 
{\em (Any(Tested(Sensitive, Resistant, Intermediate), Not-tested))}. This taxonomy describes whether for a given 
pair of a drug and a detected bug within an episode of care, or detected bug and drug (remember that the relationships are 
symmetric) there was, or there was not, an antibiogram performed, {\em Tested} and {\em Not-tested} respectively. 
If there has been a test then the possible values of the relationshipe are {\em Sensitive, Intermediate, Resistant}, which denote the
measured sensitivity of the bug on the drug. The identity relations for both \bug\ $\times$ \bug, and \trt\ $\times$ \trt,
are described by the simple prefix tree {\em (Any($=$, $\neq$))}.

\begin{table}
\begin{tabular}{|c|c|c|c|c|c|c|}
\hline 
Taxonomy   & \multicolumn{4}{c|}{Depth Statistics}& Num        & Num       \\ \cline{2-5}
       &Max  & Min        & Avg.       & Std.     & leaves     & nodes     \\ \hline \hline 
ATC    & 4   & 4          & 4.0        & 0        & 216        & 260       \\ \hline 
NewT   & 8   & 2          & 5.27       & 0.94     & 158        & 272       \\ \hline 
\end{tabular}
\caption{Statistics of the two taxonomies that we used \label{table:taxo}}
\end{table}

\begin{figure}
\caption{Histogram of \# events within episodes of care. 
}
\label{fig:histo.events}
\begin{center}
\includegraphics[width=4cm,angle=-90]{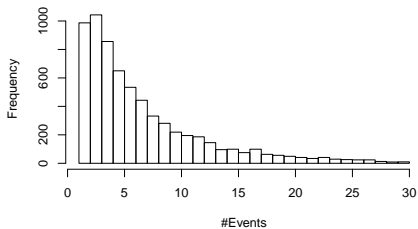}
\end{center}
\end{figure}

Some additional statistics on the available dataset. The total number of different bugs present in the dataset 
was 180. The total number of antibiotics that were tested in the different antibiograms was 67. 
The antibiograms are usually done in batches of 20 to 30 drugs for a given sample (the actual number depends on the microbe that 
is tested), which result in an antibiogram profile.
The average number of treatments and bugs detected per episode of care is 4.5 and 2.5 respectively. 
In figure~\ref{fig:histo.events} we give the histogram of the number of treatments and detected bugs per episode of care.

We investigated the performance of our method with respect to different values of the Minimum Support, MS, 
Maximum Gap, MG, and Maximum Projected Length, MPL, parameters. For MS we tested the values 300, 600, 1200, 1500,
and 1800 which correspond to 4.5\%, 9.0\%, 18\%, 22.52\%, and 27.03\% of the dataset. 
For MG we experimented with 1, 2, 3, 4, and 5. For MPL we tested with lengths
of 10, 13, 16 and 19. Whenever we were testing one of these parameters the others were set to some given value which for MS was 
10\% and 18\% of the dataset (i.e. two settings), for MG and MPL it was $\infty$ which actually means that 
there was no constraint on them.  Finally in an effort to study how RaSP scales with different dataset sizes we tested 
its performance with different subsample sizes, namely from 10\% to 100\% with a step of 10\%, in that experiment
the value of MS was fixed to 10\% of the subsample size, of MG to three and of MPL to $\infty$. The experiments were performed
in two scenarios, the {\em relationship-only} in which we did not use the taxonomies of the events so the patterns
that we were mining for where patterns of relationships between the event types, and the {\em full} scenario in which 
the event taxonomies were also included. For each experiment we present the number of frequent patterns, the 
total computational time as well as the computational time of the two stages of the algorithm, i.e. type-pattern 
mining and the subsequent step of specification of the type-patterns using the taxonomies (hierarchical mining). 
In the relationship-only scenario the computational time will be dominated by the type-pattern mining phase
since the subsequent mining phase which uses the taxonomies is limited to the small taxonomies of the relationships. 
In the full scenario the hierarchical mining step will dominate the total computational time.
We have used a Quad-Core AMD Opteron(TM) Processor 8356, 2.3Ghz and we have set the maximum heap size to 
50 GB, and as such, certain of our experiments (namely those with a small MS - $\leq\sim10\%$ - and no 
constraints) have raised an out of memory exception.

\begin{figure}
\caption{Snapshot of the Taxonomies}
\centering
\subfigure[ATC]{
\includegraphics[scale=0.18]{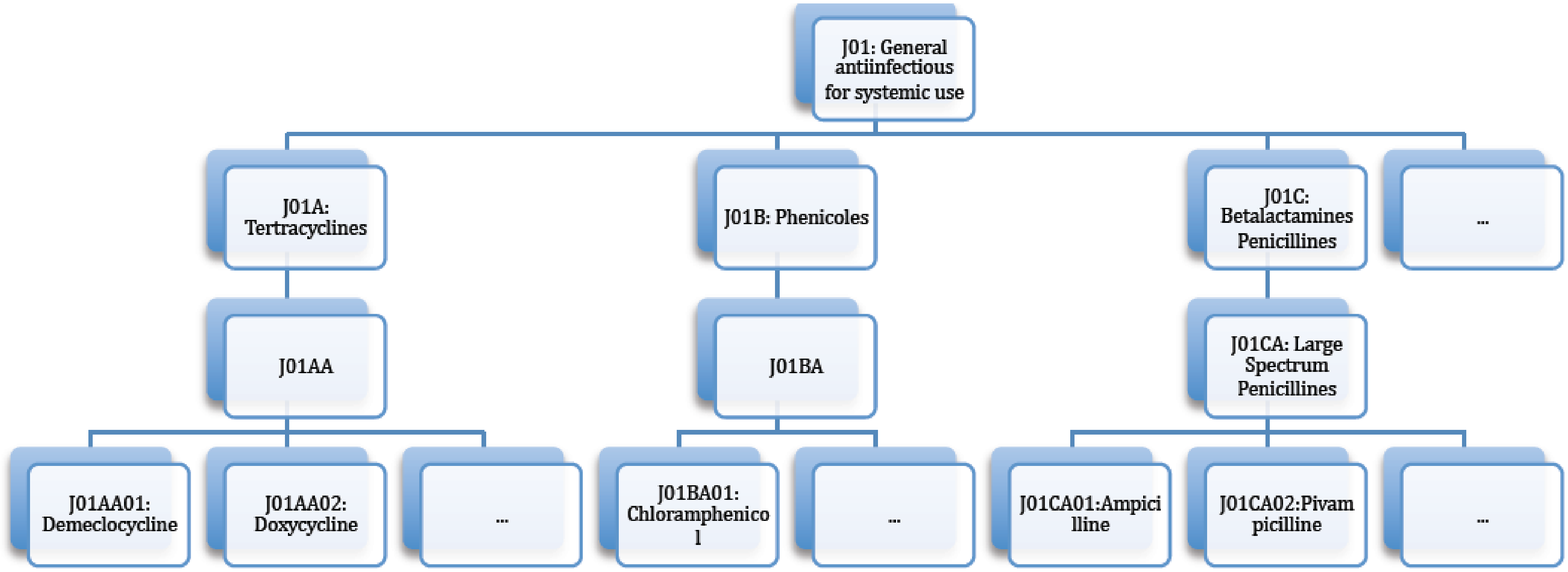}
\label{fig:ATC}
}
\subfigure[NewT]{
\includegraphics[scale=0.18]{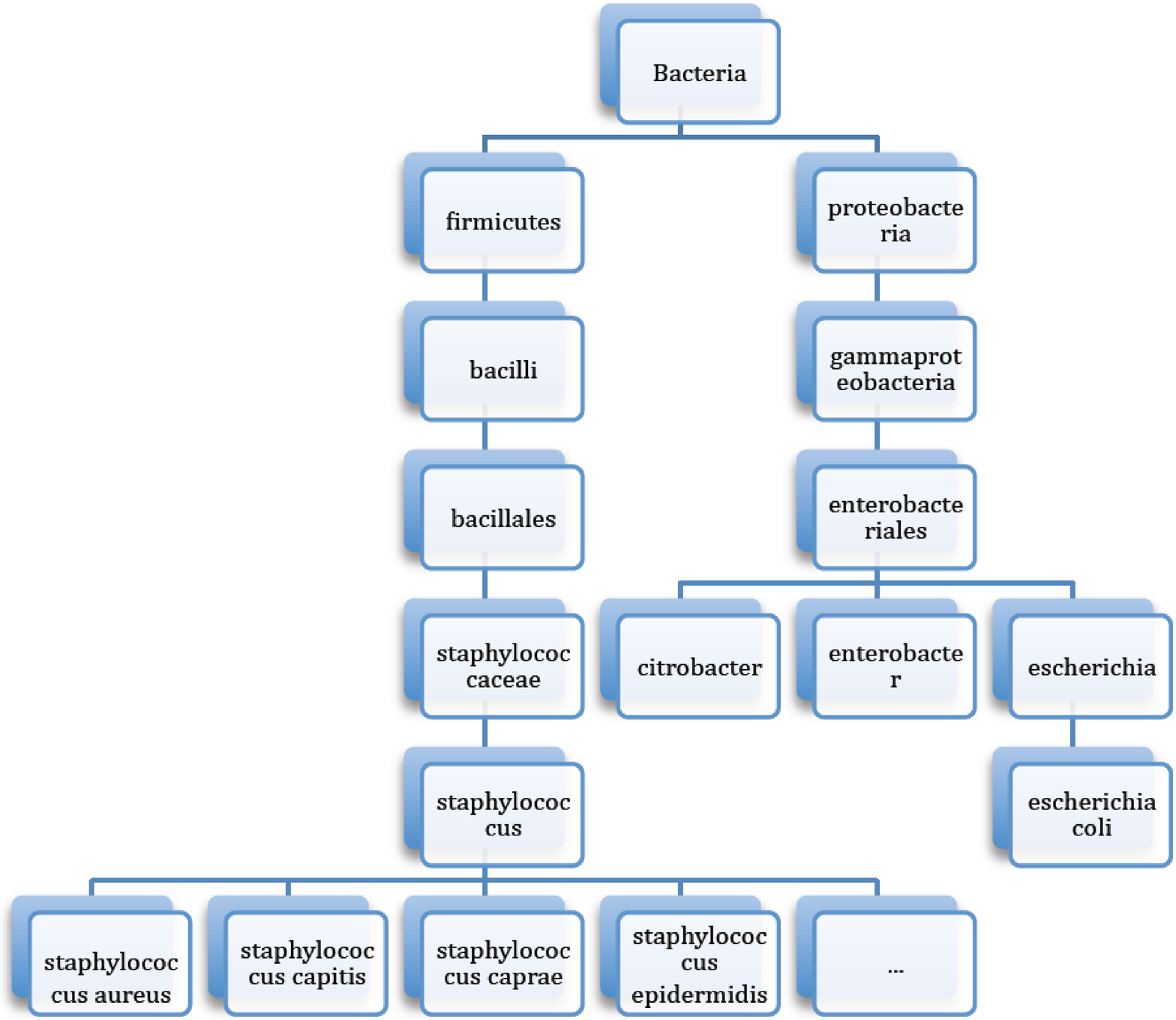}
\label{fig:NewT}
}
\label{fig:taxo}
\end{figure}

\subsection{Results}
We can see that the computational time of RaSP is affected by the value of the MG constraint with 
low values of it resulting in low execution time, top two rows of figure~\ref{fig:comptime}, something that is expected 
since the search space is considerably smaller for small MG values. For MG values smaller or equal to four the 
execution time is less than three hours, for MG=5 it is around four hours. On the same time there is a significant
increase on the number of discovered patterns which jump from less than 500 for MG=1 on the thousands for the other
values of MG. The number of patterns reduces significantly when we change MS from 10\% to 18\% for all the different
parameters we examined. The increase of the MS parameter value results into a significant decrease of computational 
time for all the different parameters we experimented with, for example for the MG parameter the decrease in computational time is almost 30-fold, for 
MPL it is smaller around five-fold. This decrease in computation time is clearly seen in the experiment
in which we test for the MS parameter, no constrains on the MPL and MG parameters,
fifth row from the top in figure~\ref{fig:comptime}. Note here that for the smallest values of the MS parameter, 
300 and 600, we got an out of memory error this is why the respective results are not included in the figure. 
In general the more constraints we add to the description of the mining problem the smaller the search space 
and the computational complexity (unlike GSP the complexity of which increases with the addition of constraints). However
less constrains do not necessarily lead to larger number of patterns,
see for example the reduction in the number of patterns with an increasing MPL, fourth row of figure~\ref{fig:comptime}.
This happens because with a less constrained definition we can get more specific patterns; this can sometimes 
lead to a decrease of the total number of patterns given that we mine for maximal patterns for each type pattern, 
because a given specific pattern can be subsumed by many more general ones but we only report the more specific 
one due to the maximal pattern property, 

Examining how the computational complexity scales with the dataset size, last row of figure~\ref{fig:comptime}, we can see that 
the computational time of the type-pattern mining increases roughly linearly with the dataset size while for the hierarchical mining part 
the behavior is more complex and irregular. The type-pattern mining part will require more time than the hierarchical mining part when 
the minimum support is set to larger values since there will be much less abstract patterns to specialize. An 
additional observation is that the discovered patterns are dominated by the 
abstract patterns, which intuitively makes sense, since the specialization of the abstract patterns will necessarily lead
to patterns that have lower support. This is made clear when we compare the number of patterns in the relationship-only scenario to the number of 
patterns in the full scenario; the number of specialized patterns that are added due to the incorporation of the full 
event taxonomies are small compared to the total number of discovered patterns.  Depending on the parameter setting the percentage 
of specialized patterns compared to the total number of frequent patterns for the full scenario ranges between 10\% and 30\%.


In table~\ref{tab:pattern.examples} we give a couple of examples of discovered patterns. The first pattern states that we have
some detected bug which is followed by some prescribed treatment to which the bug is resistant.
In the second pattern we have two subsequent treatments, which can be related in any way, treatments
that are followed by a detected bug of the gammaproteobaceria family bug which is resistant to the first treatment and it has
any relationship with the second.

\begin{figure}
\begin{tabular}{|p{3.7cm}|p{3.7cm}|}
\hline 
Relationship-only & Full \\ \hline \hline
\includegraphics[width=3.7cm]{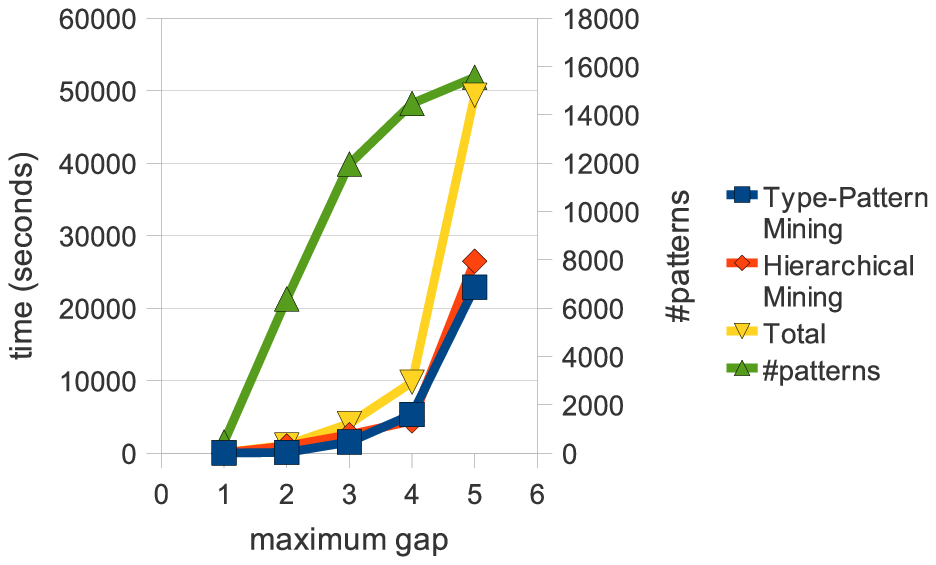} & \includegraphics[width=3.7cm]{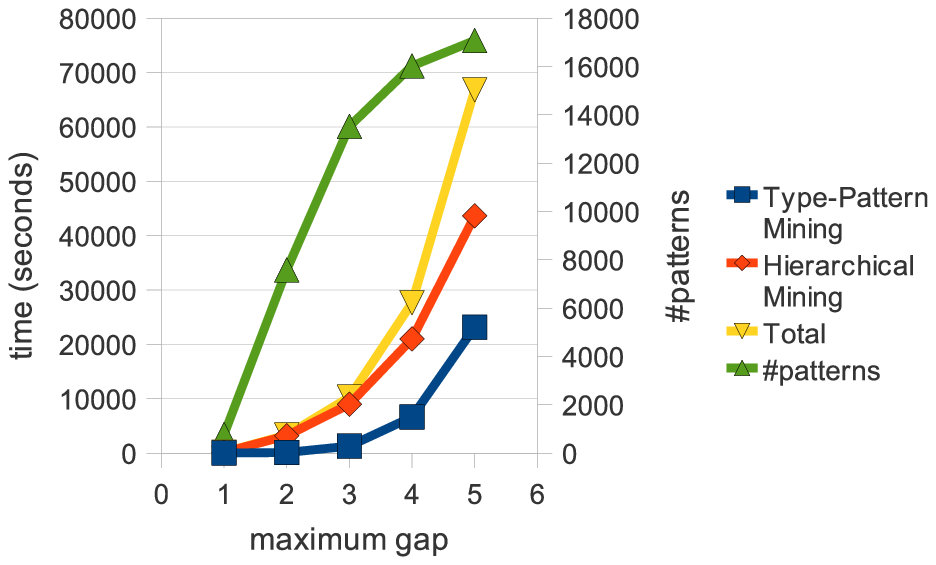} \\ \hline
\multicolumn{2}{|p{7.4cm}|}{ \small 100\% of dataset, MS=10\%, MG=1 to 5, MPL=$\infty$ }\\ \hline
\includegraphics[width=3.7cm]{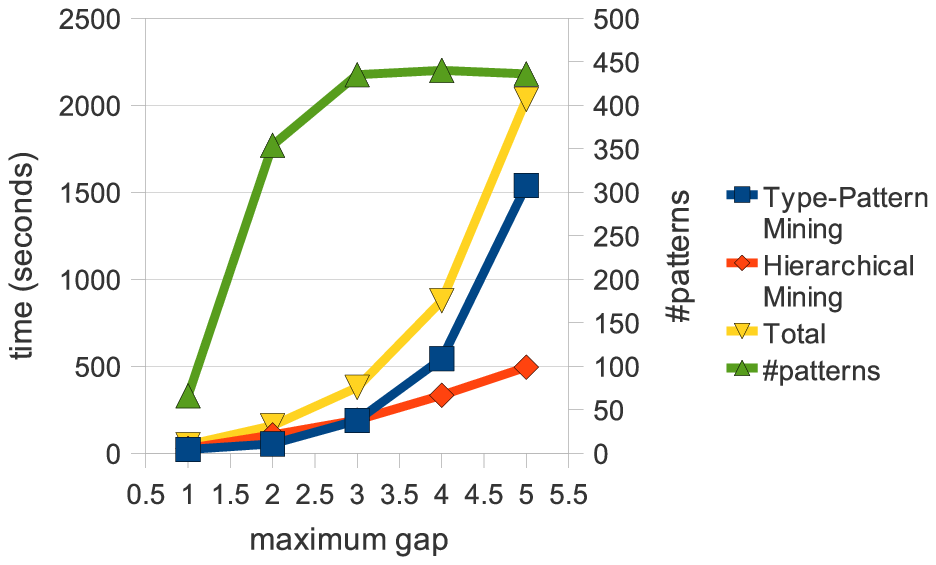} & \includegraphics[width=3.7cm]{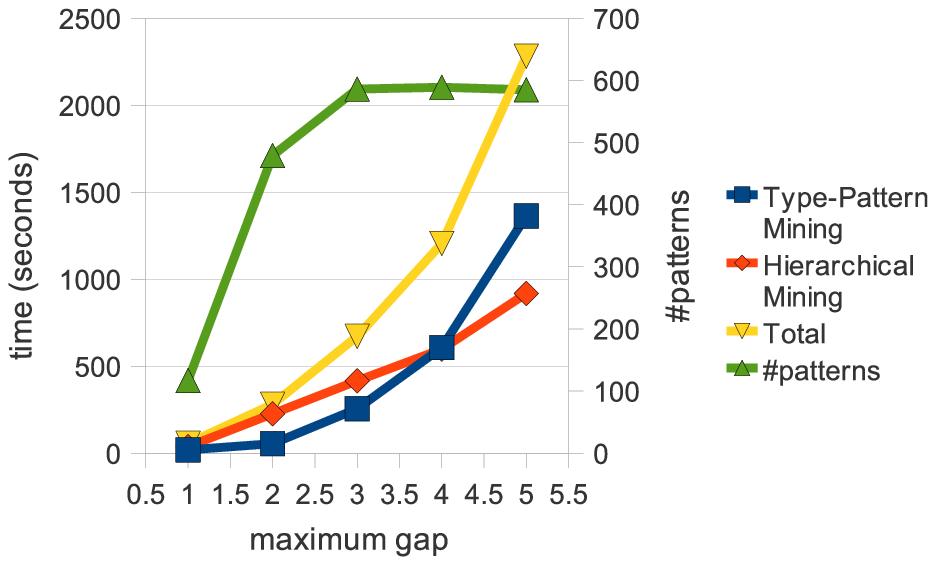}\\ \hline 
\multicolumn{2}{|p{7.4cm}|}{ \small 100\% of dataset, MS=18\%, MG=1 to 5, MPL=$\infty$ }\\ \hline
\includegraphics[width=3.7cm]{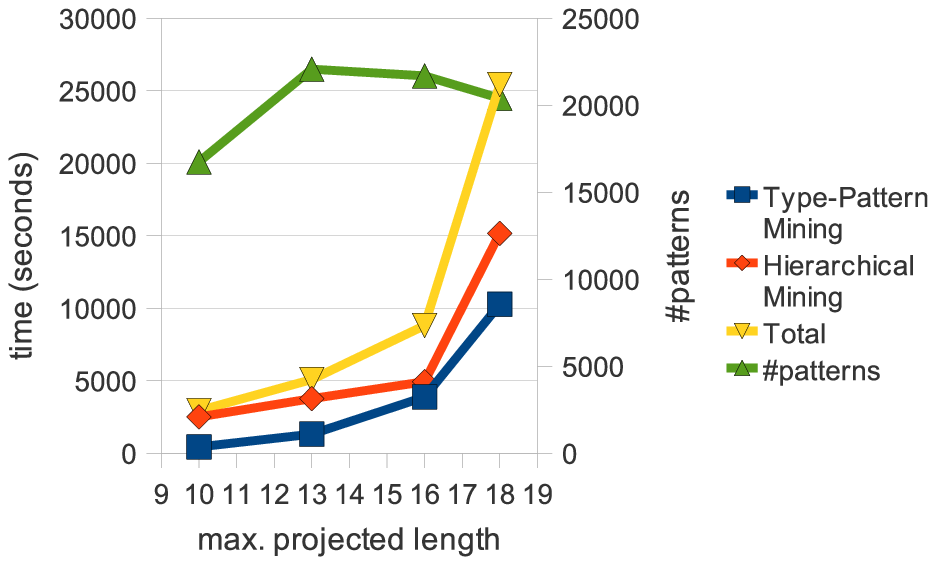} & \includegraphics[width=3.7cm]{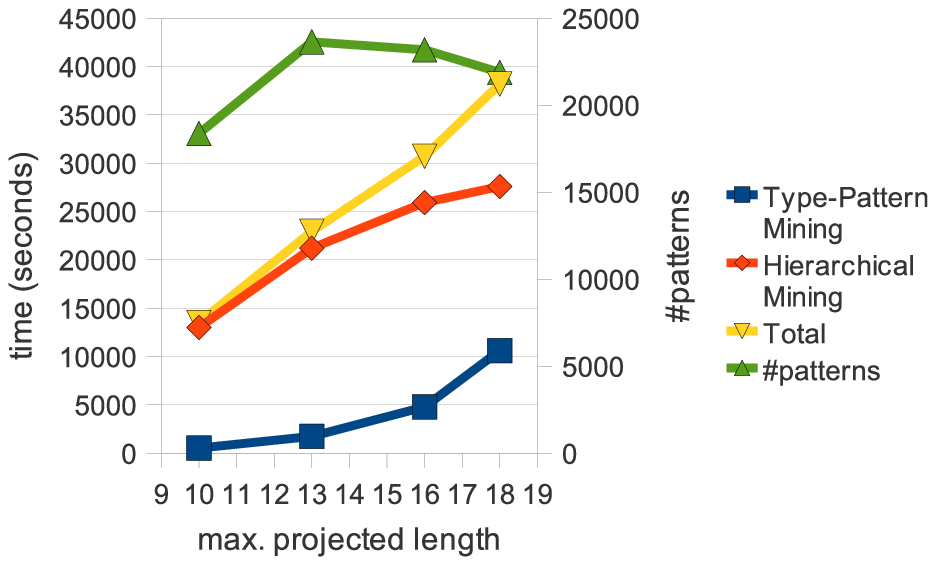}\\ \hline
\multicolumn{2}{|p{7.4cm}|}{ \small 100\% of dataset, MS=10\%, MG=$\infty$, MPL=10,13,16,19. }\\ \hline
\includegraphics[width=3.7cm]{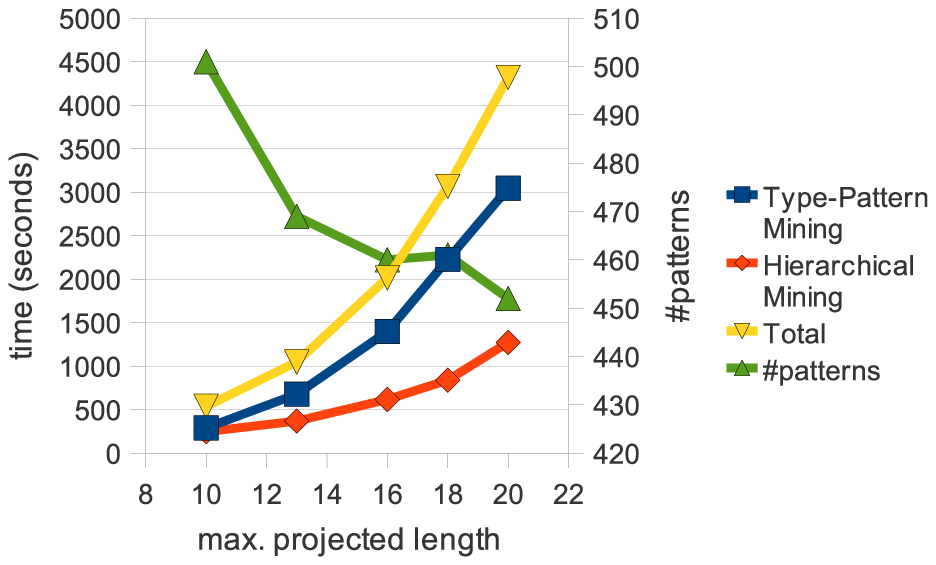} & \includegraphics[width=3.7cm]{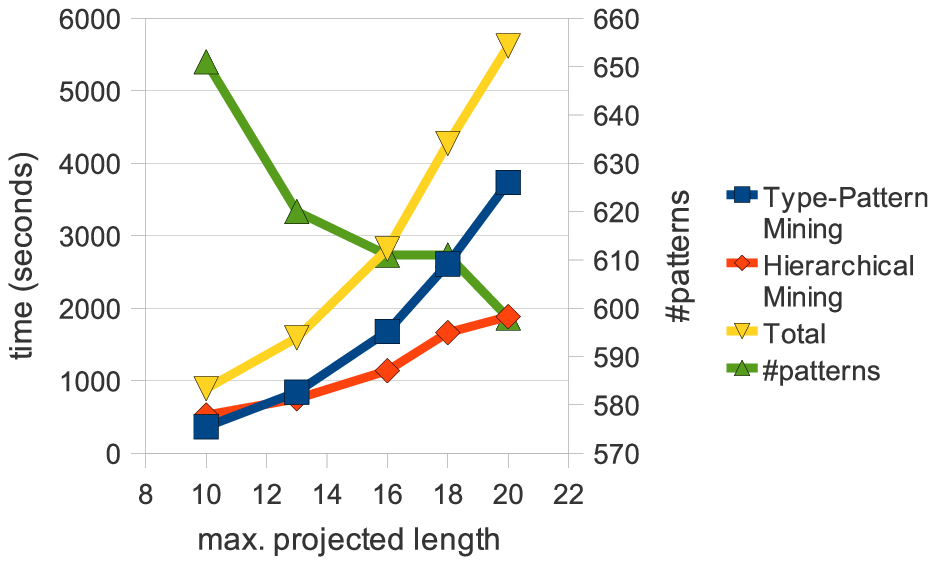}\\ \hline
\multicolumn{2}{|p{7.4cm}|}{ \small 100\% of dataset, MS=18\%, MG=$\infty$, MPL=10,13,16,19. }\\ \hline
\includegraphics[width=3.7cm]{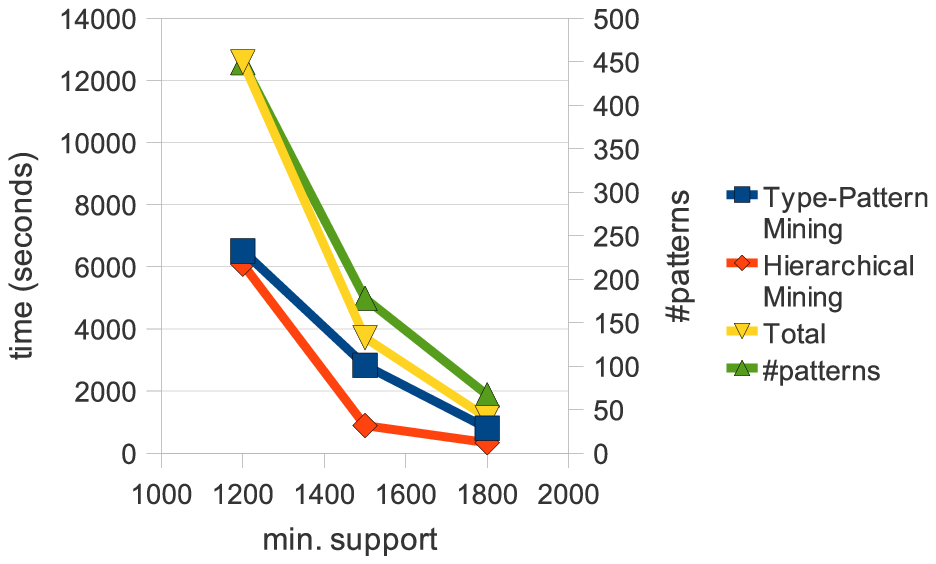} & \includegraphics[width=3.7cm]{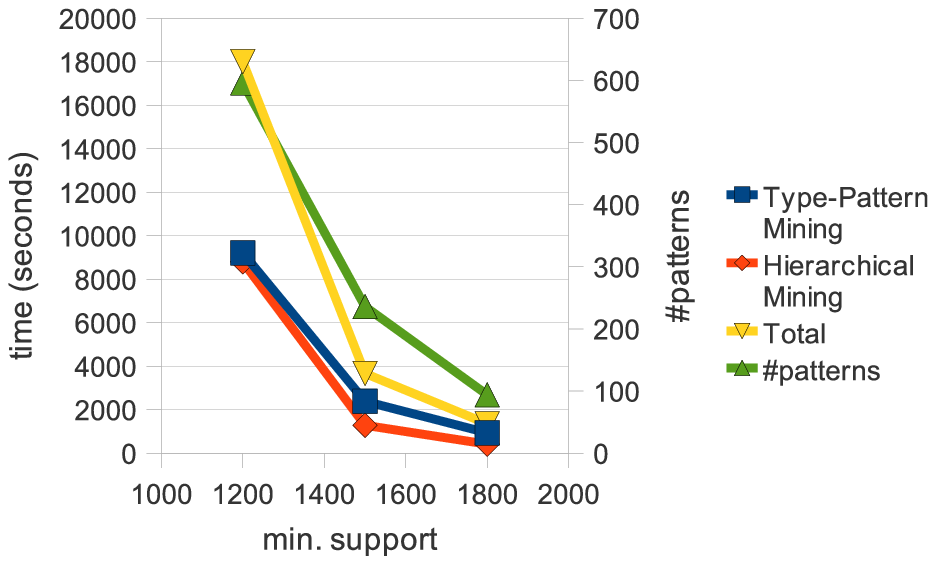}\\ \hline
\multicolumn{2}{|p{7.4cm}|}{ \small 100\% of dataset, MS=300,600, MG=$\infty$, MPL=$\infty$ }\\
\multicolumn{2}{|p{7.4cm}|}{       \small                1200,1500,1800 }\\ \hline
\includegraphics[width=3.7cm]{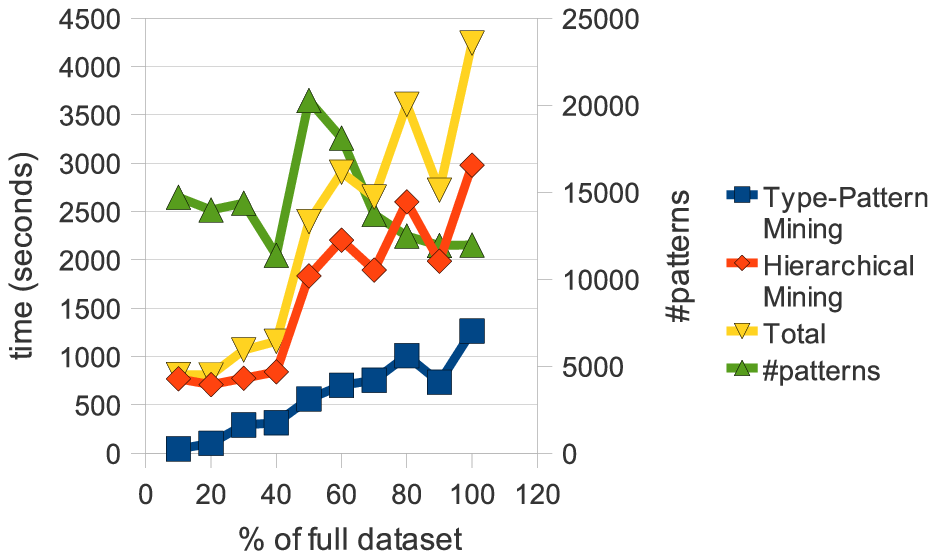} & \includegraphics[width=3.7cm]{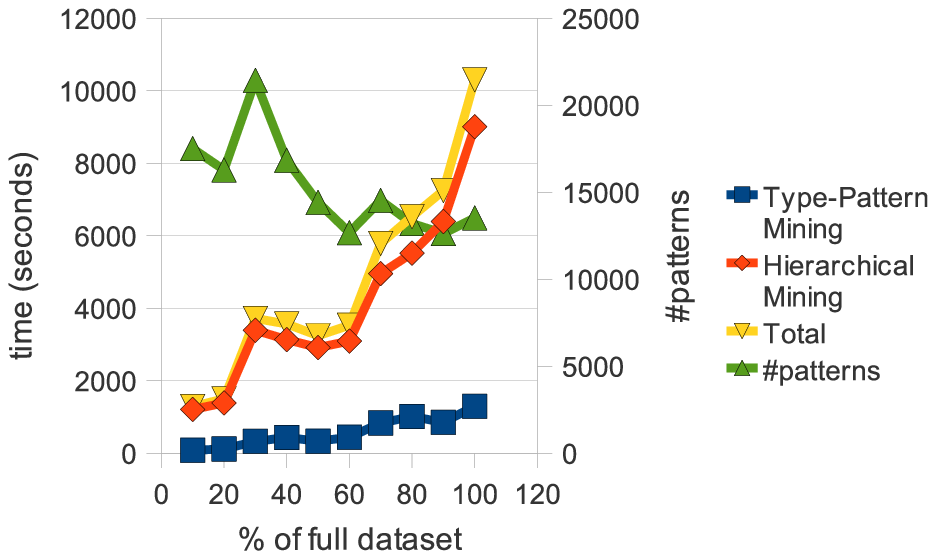}\\ \hline
\multicolumn{2}{|p{7.4cm}|}{ \small x\% of dataset, MS=10\% MG=3, MPL=$\infty$ }\\ \hline
\end{tabular}
\caption{Computational time of RaSP for various parameter settings. \label{fig:comptime}}
\end{figure}

\begin{table}
\begin{center}
\begin{tabular}{cl} \hline
Sup. \% & Pattern \\ \hline
9.20    & $B_1,\mathbf{c}("Any");T_2,\mathbf{c}("Any"),\mathbf{r}(B_1,T_2,"Resist.")$ \\ 
8.05       & $T_1,\mathbf{c}("Any");T_2,\mathbf{c}("Any"),\mathbf{r}(T_2,T_1,"Any");$ \\
           & $B_3,\mathbf{c}("gammaproteobacteria"),$\\
           & $\mathbf{r}(B_3,T_1,"Resist."),\mathbf{r}(B_3,T_2,"Any")$ \\ \hline
\end{tabular}
\end{center}
\caption{Examples of discovered patterns \label{tab:pattern.examples}}
\end{table}

\section{Related work}
\label{sec:related.work}
To perform the type-pattern mining stage of the algorithm we introduced a modified version of the GSP 
algorithm which does not rely on the detection of presence of a pattern within a sequence but instead
returns all occurences of that pattern inside the sequence.  By using occurrence itemset mining instead 
of relying on a pure Apriori-based approach, as standard GSP does, we avoid the generation of useless 
candidate patterns. The overhead of finding all occurrences of the type-patterns can often be
smaller, depending on the size of the taxonomies that are used, than generating 
and checking a huge number of candidates, most of which will not be frequent. 

Within the data mining community
the only work to our knowledge that discusses the issue of relations between the items of a sequence is 
that of \cite{DBLP:conf/kdd/MannilaT96}. The authors introduce a framework to handle unary and binary predicates 
over events, unary predicates describe properties and conditions over a single event, binary predicates describe 
relations between items, and mine over a single sequence with no taxonomies. However they only provide a sketch of the 
algorithm that allows for pattern mining over a sequence that contains binary predicates and instead focus on sequences 
that contain only unary predicates. In the domain of Inductive Logic Programming, ILP, there is a couple of works for 
frequent sequence mining over relational sequences, MineSeqLog~\cite{DBLP:conf/cinq/LeeR04}, and \cite{DBLP:journals/fuin/EspositoMBF08}.
In both mining is performed over sequences of predicates, where special predicates are introduced to describe
order; both are extensions of Apriori like algorithms in the ILP setting. Since they
are ILP approaches in principle both can model taxonomical background knowledge as 
chained lists of predicates.  However, this, aside from increasing the search space of the candidate
generation phase, will also produce frequent patterns that correspond to parts of the taxonomy relations 
without any reference to events, i.e. useless patterns. In principle this can be addressed by imposing specific constraints 
that the discovered patterns should fulfill, ~\cite{DBLP:journals/fuin/EspositoMBF08} provide the possibility of defining
constraints on the patterns (they do not discuss mining with taxonomies), however this can be quite cumbersome. In \rasp, mining 
over the taxonomies is naturally integrated in the sequence item mining. It does not lead to the generation of useless patterns 
because the construction of the patterns is done on the type-patterns which are defined over event-types; relationships 
are refined only in the next stage of the algorithm.  
MineSeqLog can be seen as an extension of frequent relational pattern mining systems, such as Warmr,~\cite{DBLP:journals/datamine/DehaspeT99},
for sequences that incorporates meaningfully the ordering operators. Its candidate generation mechanism does not have
a bias towards the generation of sequential patterns, as a result it also produces patterns that are not sequential in nature as
the example with the taxonomical patterns mentioned previously has already demonstrated. In \rasp\ this
type of bias is inherent since we first mine over type-patterns which are inherently sequential. The system proposed 
in~\cite{DBLP:journals/fuin/EspositoMBF08} does not have such a bias either however as already mentioned it provides
mechanisms through which one could declare it. In few words both approaches,~\cite{DBLP:conf/cinq/LeeR04,DBLP:journals/fuin/EspositoMBF08}, 
are more like frequent pattern mining approaches where some of the discovered frequent patterns will be of sequential 
nature. \rasp's frequent patterns are sequential by construction. Additionally we can also define relationships
over the properties of the events (our concepts in the event concept arrays) by extending the relationship concept 
arrays to include relationships over property pairs and not only over the events. Note here that
our relationships, whether between properties or events, are not limited to unification, since we can model any type
of explicit symmetric or anti-symmetric relationship. For example events difference, or events' properties difference, 
can be described as a relationship in \rasp\ but cannot be discovered in logic based approaches since for patterns of the form $a(X), a(Y)$, 
it does not necessarily hold that $X \neq Y$; modelling that in a logic based approach requires the 
incorporation of additional predicates which would again produce additional meaningless frequent 
patterns that contain only these predicates.

\section{Conclusion and future work}
\label{sec:conclusion}
We have presented \rasp\ an algorithm for mining frequent patterns from 
relationship-aware sequences in the presence of taxonomies that describe
not only the events but also the relationships between them. This is one
of the few systems that are able to do this type of mining  and unlike 
its inductive logic programming counterparts which are upgrades of relational 
frequent pattern miners to sequences it discovers naturally sequential patterns 
and copes naturally with the taxonomic information. Our approach can be easily 
generalized to structures other than sequences, 
e.g. graphs, trees, with hierarchical concepts as atoms. Given an algorithm 
that within these structures finds frequent patterns and all their occurrences 
at the root level of the taxonomy, i.e.  our type-patterns, we can then derive 
vectors from each occurrence, and proceed as before to mine the frequent patterns.

\bibliographystyle{plain}
\bibliography{rasp}

\end{document}